\documentstyle[12pt]{article}

\def \be{\begin{equation}}
\def \ee{\end{equation}}
\def \bea{\begin{eqnarray}}
\def \eea{\end{eqnarray}}
\def \l1{ \bf{\lambda_1}}
\def \l2{ \bf{\lambda_2}}

\begin{document}
{\hskip 1cm  }
\\
\hskip 6cm hep-th/9707173
\begin{center}
{\Large{\bf{ Exact Two--Point Correlation Functions of Turbulence Without 
Pressure in Three--Dimensions 
}}}
\vskip .5cm   
{\large{A. R. Rastegar${}^1$, M. R. Rahimi Tabar${}^{2,3}$ and P. Hawaii${}^1$ }}
\vskip .1cm
{\it{1) Department of Physics, Tabriz University, Tabriz, Iran
\\2) Dept. of Physics , Iran  University of Science and Technology,\\
Narmak, Tehran 16844, Iran.
\\3) Institute for Studies in Theoretical Physics and 
Mathematics
\\ Tehran P.O.Box: 19395-5746, Iran.}}
\end{center}

\vskip .5cm
\begin{abstract}
 
We investigate exact results of isotropic turbulence in  
three--dimensions when the pressure gradient is negligible.
We derive exact two--point correlation functions of density in  
three-dimensions and show that the density--density correlator
behaves as $ |{\bf {x_1 - x_2}}|^{-\alpha _3}$, 
where $\alpha_3 = 2 + \frac{\sqrt{33}}{6}$. 
It is shown that, in three--dimensions, the 
energy spectrum  $E(k)$ in the inertial range scales with  
exponent  $ 2 - \frac {\sqrt{33}}{12} \simeq 1.5212$. 
We also discuss the time scale for which our
exact results are valid for strong 3D--turbulence in the 
presence of the pressure. 
We confirm our predictions by using the recent results of
 numerical calculations and experiment.
\end{abstract} 
\vskip .5cm

PACS numbers 47.27.AK, 47.27.Jv
\newpage
{\bf 1- Introduction}

Recently, tremendous activities have started on the 
non-perturbative understanding of turbulence [1-12].
A statistical theory of turbulence has been put
forward by Kolmogorov [13], and further developed by others [15--17].
The approach is to model turbulence using stochastic
partial differential equations.  
The simplest approach to turbulence is the Kolmogorov's dimensional  
analysis, which leads to the celebrated $ k^{- 5/3}$ law for the energy spectrum. 
This is obtained by decreeing that the energy spectrum depends neither on 
the wavenumber where most of the energy resides, nor on the wavenumber
of viscous dissipation. Kolmogorov conjectured that the scaling 
exponents are universal, independent of the statistics of large--scale
fluctuation and the mechanism of the viscous damping, when
the Reynolds number is sufficiently large. In fact the idea of universality
is based on the notion of the "inertial subrange". By inertial subrange
we mean that for very large values of the Reynolds number there is a wide 
separation between the scale energy input $L$ and the typical viscous
dissipation scale $\eta$ at which viscous friction become important and 
 the energy is turned into heat.

However, recently 
it has been found that there is relation between the 
probability distribution function (PDF) of velocity and those of the  
external force [18]. This observation has been confirmed by experiments [19], 
and numerical simulations [20].

In this direction, Polyakov [5] has 
recently offered a field theoretic method to derive the 
probability distribution or density of states in (1+1)-dimensions 
in the problem of randomly driven Burgers equation [21]. 
In one dimension, turbulence without pressure
is described by Burgers equation (see also [14] concerning the relation between 
Burgers equation and KPZ--equation). 
In the limit of high Reynold's number, using 
the operator product expansion (OPE), Polyakov reduces the problem of 
computation of correlation functions in the inertial subrange, 
to the solution of a certain partial 
differential equation [22,23], see also [28], about generalization
of Polyakov's approach, to find the probablity density and 
scaling exponent of the moments of "longitudinal" velocity difference in the 
three--dimensional strong turbulence.

In this paper we consider three--dimensional 
isotropic turbulence without pressure, which is described
by Navier--Stokes equations, when the pressure gradient is negligible.
We derive the Polyakov's master equations in higher dimensions
and solve it, in the three dimensions. 
We derive the exact exponent of two-point density correlation functions
and the energy spectrum exponent. We also
 discuss the time scale for which our exact results are valid for 
strong  3D--turbulence in the presence of the pressure.
\vskip .3cm

{ \bf {2-  Turbulence Without Pressure in Three--Dimensions} }

We consider the following quantity: 
\be
\bf {e_{\bf{ \lambda}}} = \rho (\bf{x}, t) 
\exp ( {\bf {\lambda .  u ( x)}})
\ee
where $\rho$ and $\bf u$ are the density and the velocity satisfying the 
 Navier--Stokes equations:
 
\be
{\bf u}_t + ({\bf u} \cdot \nabla ) {\bf u} = \nu \nabla^2 {\bf u} - 
\frac {\nabla p}{\rho} + {\bf f}({\bf x},t)
\ee
\be
 \rho_t + \partial_{\alpha} (\rho u_{\alpha})=0
\ee
where $p$ and $\nu$ are the pressure and viscosity, respectively.
The stirring force
 $ {\bf f}({\bf x},t)$ is a Gaussian random force with the following 
 correlation:
\be
< f_\mu ({\bf{ x}},t)  f_\nu ({\bf x^{'}},t^{'})> =  k_{\mu \nu} 
({\bf {x}- { x^{'}}}) \delta (t-t^{'}) 
\ee
where $\mu, \nu = x_1, x_2, \cdots ,x_N$.

We start with the situation when $\nabla p \simeq 0$. 
This mode has a characteristic time of $T_p \simeq \frac {L}{C_s}$,
where $L$ and $C_s$ are the  
large scale of the system and the velocity of sound in the turbulent
system, respectively. The existence of this time scale means
that for the times $t < T_p$, we can ignore the pressure term in the 
Navier--Stokes equations and therefore our results are valid 
only for this time scale.

To investigate the statistical description of eqs.(2) and (3),  
 we consider the following two--point generating functional
\be 
F_2 ({\bf \lambda_1, \lambda_2, x_1, x_2}) = 
<   \rho( \bf{x_1})  \rho(\bf{x_2})    \exp (
{\bf \lambda_1 .  u ( x_1) +  \lambda_2 .  u( x_2)} )>
\ee

We write the eq.(2) and (3) in two points $\bf x_1$ and $\bf x_2$ for 
$u_1,u_2, \cdots, u_N$ and $\rho(x)$ 
and multiply the equations in $ \rho( { \bf{x_2}}) $, 
$\lambda_{1 x_1} \rho( { \bf{x_1}}) \rho( { \bf{x_2}}) $, 
$ \cdots $, $\lambda_{1 x_N} \rho( { \bf{x_1}}) \rho( { \bf{x_2}}) $ 
and $\rho( { \bf{x_1}})$, 
$\lambda_{2 x_1} \rho( { \bf{x_1}}) \rho( { \bf{x_2}})$, $\cdots$,
and $\lambda_{2 x_N} \rho( { \bf{x_1}}) \rho( { \bf{x_2}})$, respectively.

We add the equations and multiply the result to
$ \exp ({ \bf \lambda_1} . { \bf u (\bf x_1)} + {\bf \lambda_2 }. {\bf u({\bf x_2})} )$
and make averaging with respect to external random force, therefore we find: 

\be
\partial _t F_2 + \sum _{\{i=1,2\} \mu = x_1, \cdots, x_N}   \frac { \partial }
{ \partial \lambda_{i,\mu}}  \partial_{\mu_i} F_2 =  \sum_{\{i=1,2 \} \mu = x_1, \cdots, x_N}
 C_{i, \mu} + D_2
\ee
where  $C_{i, \mu}$ and $ D_2 $ are,

\be
\sum_{\{i=1,2 \}\mu = x_1, \cdots, x_N}  C_{i,\mu} = 
\sum_{\{i,j=1,2 \}\mu, \nu = x_1, \cdots, x_N}  
\lambda_{i, \mu} \lambda_{j, \nu} k_{\mu \nu} ({\bf x_i} - {\bf x_j}) F_2
\ee
and 
\bea 
D_2 = < \nu \rho( { \bf{x_1}}) \rho( { \bf{x_2}}) [ 
{\bf {\lambda_1}} \cdot \nabla^2 {\bf{ u }}({\bf {x_1}}) + {\bf {\lambda_2}} 
\cdot \nabla^2 {\bf {u}( {x_2})} ]      
\exp {( {\bf {\lambda_1}} \cdot {\bf{ u }}({\bf {x_1}}) + 
{\bf {\lambda_2}} \cdot {\bf {u}( {x_2})} )}>
\eea
where we have used the Novikov's theorem. $D_2$ is known as the anomaly 
term [5]. 

Now we consider the anomaly term in the limit  
of small $\nu$ or high Reynold's numbers. It is noted that 
this term can not be written in terms of $F_2$.
To find its structure we consider the symmetries of the 
basic equations. The basic equations are Galilean invariant and also are
invariant under the rescaling of density as  $\rho \rightarrow \alpha \rho$.
In the other hand, final expression for $D_2$ must contain the  
ultraviolet finite operators $ \nabla {\bf u}$, $\rho $
and $ e^{\bf{\lambda \cdot u}}$.
The only finite combination satisfying the rescaling  
$\rho \rightarrow \alpha \rho$ is $\rho e^{{\bf {\lambda \cdot u}}}$ (see [24] for 
more detail). Therefore $D_2 $ has the following form:
\be
D_2 = a F_2
\ee
where $a$ is generally a function of $\lambda_1$ and $\lambda_2$.

Therefore in the steady state we find the following equation for $F_2$:
\bea
&&\sum_{\{i=1,2\} \mu= x_1, \cdots, x_N}  \frac {\partial}{\partial \lambda_{i,\mu}}
\partial _{\mu_i } F_2 
 \nonumber \\ && -\sum_{\{i,j=1,2\} \mu,\nu = x_1,\cdots , x_N} 
\lambda_{i, \mu} \lambda_{j, \nu} k_{\mu \nu} ({\bf x_i}- {\bf x_j}) F_2
= a F_2 .
\eea
Also in  this paper, we suppose that $k_{\mu \nu}$ has the following form:
\be
k_{\mu \nu} ({\bf x_i} - {\bf x_j}) = k(0) [1- 
\frac {| {\bf x_i} - {\bf x_j}|^2}{2 L^2} \delta_{\mu,\nu} -
\frac {({\bf x_i} - {\bf x_j})_\mu ( {\bf x_i} -  {\bf x_j})_\nu}{L^2} ]
\ee
with $k(0), L=1$. 

Now let us consider the eq.(10) in three dimensions. We change the variables as:
$\bf x_\pm = \bf x_1 \pm \bf x_2 $, $ \bf \lambda_+ = \bf \lambda_1 + \bf \lambda_2$ 
and 
$\bf \lambda_-  = \frac { {\bf \lambda_1} - {\bf \lambda_2} }{2}$ and 
and consider the spherical coordinates, so that 
$ x_- : (r,\theta, \varphi)$ and $ \lambda_-: (\rho', \theta ', \varphi')$.
Direct calculation shows that: 
\bea
\sum_{i=1, \mu= x,y,z} ^3 \frac {\partial}{\partial \lambda_{i,\mu}}
\partial _{\mu_i  } &=&
\sum_{\mu= x,y,z}  \frac {\partial}{\partial \lambda_{- \mu}}
\partial _{\mu_- } =
cos {\gamma} \partial_r \partial_\rho' \nonumber \\ &+& 
\frac { sin{\theta} cos {\theta'}
cos (\varphi - \varphi') -cos{\theta} sin {\theta'}}{\rho'} \partial_r \partial_{\theta'} 
+
\frac {sin{\theta} sin(\varphi - \varphi') } {{\rho'} sin{\theta'}} \partial_r \partial_{\varphi'} 
\nonumber \\ &+&
\frac {sin{\theta'} cos {\theta}
cos (\varphi - \varphi') - cos{\theta'} sin {\theta}}{r} \partial_\rho' \partial_{\theta} 
 +
\frac {cos{\theta} sin(\varphi - \varphi') }{r \rho' sin {\theta'}} \partial_\theta  \partial_{\varphi'} 
\nonumber \\ &-&
\frac {cos{\theta'} sin(\varphi - \varphi') }{r \rho' sin {\theta}} \partial_{\theta'} \partial_{\varphi}
+
\frac {cos (\varphi - \varphi') }{r \rho' sin {\theta} sin {\theta'}} \partial_{\varphi}  \partial_{\varphi'} 
\nonumber \\ &+&
\frac { cos{\theta} cos {\theta'}
cos (\varphi - \varphi') + sin{\theta} sin{\theta'}}{r \rho'} \partial_{\theta} \partial_{\theta'} 
\nonumber \\ &-&
\frac {sin{\theta'} sin(\varphi - \varphi') }{r sin {\theta}} \partial_\rho' \partial_{\varphi} 
\eea
and 
\bea
&&\sum_{\{i,j=1,2\} \mu,\nu = x,y,z}  
\lambda_{i, \mu} \lambda_{j, \nu} k_{\mu \nu} (\bf{ x_i}- \bf{ x_j}) \nonumber \\ &=&
[r^2 \rho'^2 + 2(x_- \lambda_{-x} + y_- \lambda_{-y} + z_- \lambda_{-z})^2] 
=  r^2 \rho'^2 (1+2 cos^2 {\gamma})
\eea
where $ cos{\gamma} =
 cos{\theta} cos {\theta'} + sin {\theta} sin{\theta'} cos(\varphi - \varphi')$.  


Now using the eqs.(12,13) for isotropic turbulence we obtain: 

\bea
&&[ s \partial_r \partial_\rho' - \frac{s(1-s^2)}{r \rho'} \partial^2 _s 
+ \frac {1+s^2}{r \rho'} \partial_s  \nonumber \\ &&
+ \frac {1-s^2}{\rho'} \partial_r \partial_s 
+ \frac {1-s^2}{r} \partial_\rho' \partial_s 
- r^2 \rho'^2 (1 + 2 s^2) ] F_2 = a(\rho') F_2
\eea
where $cos \gamma = s$.
The $\rho'$ dependence of the $a(\rho')$ anomaly must be chosen to conform the 
scaling and can be different depending on the scaling properties of the 
force correlation functions. In general, in the case of isotropic 
turbulence, stirring correlation
function behaves as $k_{\mu,\nu} \sim 1 - r^{\eta}$, where in our  
case we have $\eta=2$. Therefore, $a$ must depend on $\rho'$ as follows 
$a(\rho') = a_0 {\rho'}^{\sigma} $, where $\sigma = \frac {2 - \eta} {1+ \eta}$. 
It is evident that for our case $a$ is independent of $\rho'$.
Let us consider the universal scaling invariant solution of eq.(33) in the
following form:
\be
F_2 (\rho' , r) = g(r) F(\rho' r^{\delta}) \hskip 2cm g(r) = r^{-\alpha_3}
\ee
where $\delta = \frac {\eta + 1}{3}$, and $\alpha_3$ is 
the exponent of two-point correlation
functions of density and also using the eq.(11) we find $\delta = 1$.

We substitute eq.(15) in eq.(14), and find the following relation for $F(\rho' r)$: 
\bea
&&[-\frac{\alpha_3 s}{r} \partial_\rho' + s \partial_\rho' \partial_r
-\frac{s (1-s^2)}{r \rho'} \partial_s ^2 + \frac{1+s^2}{ r \rho'} \partial_s
\nonumber \\ &&
- \alpha_3 \frac{1-s^2}{r \rho'} \partial_s 
+ \frac {1-s^2}{\rho'} \partial_s \partial_r
+ \frac {1-s^2}{r} \partial_s \partial_\rho'
- r^2 \rho'^2 (1 + 2 s^2)] F(\rho' r) = a_0 F(\rho' r)
\eea
 
 Since the two point density correlators exists, in the limit 
of $\rho' \rightarrow 0$, $ F( \rho' r)$ tends to a constant and thus 
we have to look for the solution of $F$ among the family 
of positive, finite and normalizable solution of eq.(16).
In the other hands, taking the  Laplace transformation of the above 
equation, one can show that, to consider
physical solution, so that $< {\bf u}({\bf x_1}) - {\bf u}({\bf x_2})> = 0$, 
we have to consider 
the case $a_0 = 0$. However for different types of 
correlation for the stirring force, e.g.  $k_{\mu,\nu} \sim 1 - r^{\eta}$,
with $\eta \neq 2$, we have to include the $a_0$ [23].

Now, we propose the following Ansatz for $F(\rho' r)$, with $z=\rho' r$:
\be
F(z,s) = e^{ z^\gamma f(s)}
\ee
Using the eq.(16), we find $\gamma=3/2$ and $f(s)$ satisfy the 
following equations:
 \be
\left \{ \begin {array} {ll}  
\frac {9}{4} s f(s) ^2 + 3 f(s) f'(s) (1-s^2) + f'(s) ^2 (-s + s^3) = 
(1+2s^2) &
\\
-s(1-s^2) f(s) ^{''} + [(4+\alpha_3)- (2+\alpha_3) s^2] f^{'}(s) 
+ (\frac{9}{4} + \frac {3}{2} \alpha_3 ) s f(s) = 0
\end{array} \right.
\ee

also from eq.(18-a), one can derive the 
following initial conditions for $f(s)$,
\be
f(1) = \frac {2}{\sqrt 3}  \hskip 2cm 
f^{'}(1)= \frac { \sqrt 3 + 
 \sqrt 11}{4}
\ee
 
 It is interesting to note that the equation for $f(s)$ (i.e. eq.(18-a)) is the same  
 as equation which is found in the instanton approach [8]. 
 
The function $f(s)$ has the following expansion around $s=1$:
\be
f(s) = \frac {2} {\sqrt 3} + \frac { \sqrt{3} + \sqrt {11}}{4} (s-1)
+ \frac {5 \sqrt {33} - 61}{32 (3 \sqrt{3} - 2 \sqrt{11})} (s-1)^2 + \cdots
\ee

Now using the boundary conditions on $f(s)$ (i.e. eq(19)), and  
positivity of the probability distribution function we find: 
\be
\alpha_3 =  \frac{12 + \sqrt{33}}{6} \simeq 2.9574
\ee

Noting the fact that $\rho'$ has dimension $-1$, we can find the 
following scaling relation for the density of the energy $ \epsilon(x)$,
\be 
\epsilon (\alpha x) = \alpha^{\Delta} \epsilon (x)
\ee
where $ \Delta = 1 - \frac{\sqrt{33}}{12}$, and therefore we can 
determine the behavior of the energy spectrum exactly as:
\be
E(k) \sim k^{- \beta}, \hskip 2cm  \beta= 2 - \frac{\sqrt{33}}{12} \simeq
1.52128
\ee

This behaviour of energy spectrum 
is known as the non--Kolmogorov power laws which 
has been observed experimentally [25,27] and also in 
the numerical simulations [25,26].

The numerical calculations have been done in [25,26], where they have 
used the Wiener--Hermit expansion. They have shown  
that the energy spectrum behaves as  $E(k) \sim k^{-1.521}$
for systems without boundaries (i.e. free turbulence) and also for a finite  
system this spectrum is not stable. In [25,26] it has been shown that 
in the inertial subrange for a finite system energy spectrum starts with
slope $-1.521$, and after a moderate time which is less than the 
characteristic time $T_c \simeq \frac {L}{u_{rms}}$ ( where
$L$ and $u_{rms}$ are the large scale of the system and rms value
of initial velocity fluctuation, respectively) the equilibrium is attained  
and has transformed to $- 5/3 $. In other words the $-5/3$'s law is the stable 
algebraic spectrum for the Navier--Stokes equations after a time of order
$T_c$.

The experimental results (reported by Wissler [25,27]) shows that the 
non--Kolmogorov spectrum has been observed also experimentally only 
for moderate times less than $T_c$. It is noted that in general
for a turbulent flow $u_{rms} \leq C_s$ and therefore the pressure 
time scale has the property that $T_p \leq T_c$.

Finally we can derive the PDF for the velocity difference and show 
that it tails as $ e^{-\alpha u^3}$ in the limit 
$|u| \rightarrow + \infty $, which is in agreement with other approaches [18]
for three--dimensional turbulence. 
\\
{\bf Acknowledgements:} We would like to thank B. Davoudi
, J. Davoudi, R. Ejtehadi, M. Jafarizade, A. Langari 
M.R. Mohayaee and S. Rouhani for 
useful discussions and D.D. Tskhakaya for important comments.


\newpage 
\begin{thebibliography}{99} 
\bibitem{1} A. Polyakov, Nucl. Phys. {\bf B396} 367 (1993)
\bibitem{2} A. Migdal, Int. J. Mod. Phys. {\bf A9} 1197 (1994)
\bibitem{3} V. L'vov and I. Procaccia, chao-dyn/9502010 
\bibitem{4} J. Bouchaud, M. Mezard and G. Parisi, Phys. rev. E {\bf 52} 3656 (1995)
\bibitem{5} A. Polyakov, Phys. Rev. E {\bf 52}, 6183 (1995)
\bibitem{6} E. Balkovsky, G. Falkovich, I. Kolokolov and V. Lebedov, 
JETP Lett. {\bf 61}, 1012 (1995)
\bibitem{7} M. Chertkov, Phys. Rev. E {\bf 55}, 2722 (1997), chao-dyn/9606011
\bibitem{8} V. Gurarie and A. Migdal, Phys. Rev. E {\bf 54} 4908 (1996)
\bibitem{9} J. P. Bouchaud, M. Mezard, cond-mat/9607006
\bibitem{10} K. Gawedzki, chao-dyn/9610003
\bibitem{11} D. Antonov, hep-th/9612005
\bibitem{12} P. Tomassini, chao-dyn/9706013
\bibitem{13} A. N. Kolmogorov, C. R. Acad. Sic. USSR {\bf 30}, 301 (1941)
\bibitem{14} M. Kardar, G. Parisi and Y. Zhang, Phys. Rev. Lett. 56, 889 (1986)
\bibitem{15} A. S. Monin and A. M. Yaglom, " Statistical Fluid Mechanics, 
(MIT Press, Cambrige, MA, 1975)
\bibitem{16} C. DeDominicis and P. C. Martin, Phys. Rev. A {\bf 19} 419 (1979)
\bibitem{17} V. Yakhot and S. A. Orszag, J. Sci. Comp. {\bf 1}3 (1986) 
\bibitem{18} G. Falkovich and V. Lebedev, chao-dyn/9708002
\bibitem{19} A. Noullez, G. Wallaec, W.Lempert, R.B. Miles
and U. Frisch, J. Fluid Mech. {\bf 339} 287 (1997)
\bibitem{20} R. H. Kraichnan and Y. Kimura, in "Progress
in Turbulence Research" (1994), pp-19
\bibitem{21} J. M. Burgers, The Nonliner Diffusion Equation (Reidel, Boston, 1974)
\bibitem{22} M. R. Rahimi Tabar, S. Rouhani and B. Davoudi, Phy. Lett. A {\bf 212}
60 (1996)
\bibitem{23} S. Boldyrev, Phys. Rev. E {\bf 55}, 6907 (1997), hep-th/9610080
\bibitem{24} S. Boldyrev, hep-th/9707255
\bibitem{25} T. Funaki and W. A. Woyczynski "Nonlinear Stochastic PDEs"
springers (1995), pp--293-312
\bibitem{26} T. C. Chung, Ph.D. Thesis, university of California at Los Angeles (1991)
\bibitem{27} J. B. Wissler, presentation to meeting of the Div. of 
Fl. Dynamics, Am. Phys. Soc. Nov (1993)
\bibitem{28} V. Yakhot, Princeton Univ. Preprint, chao-dyn/9708016

\end{thebibliography}
\end{document}